\documentclass[conference]{IEEEtran}
\IEEEoverridecommandlockouts
\usepackage{cite}
\usepackage{amsmath,amssymb,amsfonts}
\usepackage{array}
\usepackage{algorithmic}
\usepackage{graphicx}
\usepackage{multirow}
\usepackage{textcomp}
\usepackage{xcolor}
\usepackage{hyperref}
\usepackage{subfiles}
\usepackage{dblfloatfix}
\usepackage{xr}
\usepackage{subfigure}

\renewcommand{\thefootnote}{\fnsymbol{footnote}}

\def\BibTeX{{\rm B\kern-.05em{\sc i\kern-.025em b}\kern-.08em
    T\kern-.1667em\lower.7ex\hbox{E}\kern-.125emX}}
    
\begin{document}

\title{ChaRRNets: Channel Robust Representation Networks for RF Fingerprinting
\thanks{This research was developed with funding from the Defense Advanced Research Projects
Agency (DARPA). The views, opinions and/or findings expressed are those of the author 
and should not be interpreted as representing the official views or policies of the 
Department of Defense or the U.S. Government.}
\thanks{Distribution Statement `A' (Approved for Public Release, Distribution Unlimited).}
}

\author{\IEEEauthorblockN{Carter N. Brown\textsuperscript{\textsection} }
\IEEEauthorblockA{\textit{} \\
\textit{Expedition Technology}\\
Herndon, VA, USA \\
carter.brown@exptechinc.com}
\and
\IEEEauthorblockN{Enrico Mattei\textsuperscript{\textsection} }
\IEEEauthorblockA{\textit{} \\
\textit{Expedition Technology}\\
Herndon, VA, USA \\
enrico.mattei@exptechinc.com}
\and
\IEEEauthorblockN{Andrew Draganov\textsuperscript{\textsection}}
\IEEEauthorblockA{\textit{} \\
\textit{Expedition Technology}\\
Herndon, VA, USA \\
andrew.draganov@exptechinc.com}

}

\maketitle

\begingroup\renewcommand\thefootnote{\textsection}
\footnotetext{Equal contribution}
\endgroup

\begin{abstract}
We present complex-valued Convolutional Neural Networks (CNNs) for RF
fingerprinting that go beyond translation invariance and appropriately account for
the inductive bias with respect to multipath propagation channels, a phenomenon that is
specific to the fields of wireless signal processing and communications.
We focus on the problem of fingerprinting
wireless IoT devices in-the-wild using Deep Learning (DL) techniques. Under
these real-world conditions, the multipath environments represented in the train
and test sets will be different. These differences are due to the physics
governing the propagation of wireless signals, as well as the limitations of
practical data collection campaigns.

Our approach follows a group-theoretic framework, leverages prior work on DL
on manifold-valued data, and extends this prior work to the wireless signal
processing domain. We introduce the Lie group of transformations that a signal
experiences under the multipath propagation model and define operations that
are equivariant and invariant to the frequency response of a Finite Impulse
Response (FIR) filter to build a ChaRRNet. We present results using synthetic and
real-world datasets, and we benchmark against a strong baseline model, that show
the efficacy of our approach. Our results provide evidence of the benefits of
incorporating appropriate wireless domain biases into DL models. We hope to
spur new work in the area of robust RF machine learning, as the 5G revolution
increases demand for enhanced security mechanisms.

\end{abstract}


\maketitle

\section{INTRODUCTION}
\label{sec:introduction}

\IEEEPARstart{F}{eedforward} neural networks are not inherently comprised of 
operations robust to transformations of the network's input. A popular way to
alleviate this problem is to increase the effective size of the training set by
presenting transformed examples to the network during training, a process known
as data augmentation.
An alternate solution is to design networks with inherent
robustness to known transformations, e.g., translation. In classification tasks,
such as image recognition and segmentation, extracting features that are
translation equivariant, i.e. translating the input results in a translated
version of the latent output, is of great importance as it produces models with better
parameter efficiency. Another important property in classification tasks is that
of translation invariance, i.e., the output features remain the same regardless
of translations of the input. Translation invariant models are both parameter
and data efficient. The combination of convolution and max pooling make CNNs
approximately translation invariant (convolution is translation equivariant,
while max pooling is approximately invariant to small translations).
Designing neural networks that exhibit the desired invariances
leads to faster learning and better generalization under different train/test
data distributions--a principal tenet of System 2 processing, as outlined by
Goyal and Bengio \cite{goyal_inductive_2020}.

Since the conception of CNNs \cite{lecun_cnn_1998},
subsequent work has continued to seek models that are equivariant and invariant to other groups of transformations.
%The work of Mallat \cite{mallat_group_2012}, Bruna and Mallat
%\cite{bruna_invariant_2012}, and Sifre and Mallat \cite{sifre_rotation_2013}
%introduced rotation, scaling, and deformation invariant feature extractors based
%on wavelet scattering networks by cascading wavelet transform convolutions and
%modulus and averaging operators. 
%More recently, CNNs that are
Recently, CNNs that are equivariant to different symmetry groups have been introduced by Cohen et al.
\cite{cohen_group_2016, cohen_convolutional_2017, cohen_general_2020}. These
networks require that the convolution is performed jointly over both the space and the
group, increasing the computational burden. Cheng et al.
\cite{cheng_rotdcf_2018} proposed a rotation equivariant CNN using decomposed
steerable filters. Under this framework, only the filter expansion coefficients
are learned, which significantly reduces the computational burden involved in
computing the group convolution. In a similar fashion, Zhu et al.
\cite{zhu_scale-equivariant_2019} proposed a scale equivariant CNN with
guaranteed representation stability under input deformations.

The theory of equivariant CNNs has thus far been primarily focused on real-valued data
with applications to computer vision. Wireless signals, such as those used in
communications and RADAR, however, are complex-valued. Real-valued networks that
treat complex-valued signals as two independent real-valued channels fail to
exploit the relationship between the I and Q components of the signal.
Furthermore, we seek CNNs that are equivariant/invariant to different transformations
than traditional CNN requirements due to the physics of the problem.
Chakraborty et al. \cite{chakraborty_surreal_2020}
introduced CNNs for complex-valued data that are invariant to non-zero scaling
and planar rotations, which is the group that acts transitively on the complex
plane. This CNN uses the polar form of the complex numbers and identifies the
non-zero complex plane with the product space of the positive real numbers and
the rotation Lie group, $SO(2)$. The standard Euclidean convolution is replaced
by a weighted Frechet Mean (wFM) \cite{frechet_mean} of points in this product
space, which can be shown to be equivariant to complex scaling. To make the CNN
invariant, they introduced a second convolutional layer, which operates on the wFM
convolution output, with the desired invariance property. This model was tested
on the tasks of synthetic aperture radar (SAR) target recognition 
and modulation recognition of communications signals.
The results showed significant improvements in parameter and data efficiency on both
the MSTAR and RadioML datasets \cite{mstar_dataset, oshea_ota_dl}.
Note that a complex scaling corresponds to a
1-tap channel. Although 1-tap channels model Line-of-Sight (LOS) propagation,
robustness to them is not sufficient for the space of all multipath propagation environments.

In this paper we propose ChaRRNets as a kind of CNN architecture for robust RF fingerprinting under
multipath propagation channels. We empirically show increased representation
stability, relative to a strong baseline model, when the channel realizations
represented in the train and test sets are drawn from different statistical
models, i.e. ChaRRNets produce better out-of-distribution generalization.
Our main contributions are as follows:

\begin{itemize}

\item We extend the work presented in \cite{chakraborty_surreal_2020}, by
formulating the problem in the frequency domain, and we introduce
convolutional NN layers that are equivariant and invariant
to a frequency response acting upon the transmitted signal's spectrum.

\item We introduce ChaRRNets: a kind of CNN architecture for multipath-robust RF fingerprinting.
Up to a boundary effect that is a function of the length of the channel's impulse
response, our network architecture exhibits the desired
stability. That is, the network's output features with and without multipath
remain approximately the same. 

\end{itemize}

%\section{Related Work}
%\label{sec:related_work}
%\subfile{sections/2_related_work}

\section{Methods and Procedures}
\label{sec:methods_and_procedures}

\subsection{RF Fingerprints}
\label{subsec:fingerprinting}

In a typical RF transmit chain, the I and Q components go through
independent low-pass filters or Digital-to-Analog Converters (DACs) and are then
quadrature modulated and amplified before being transmitted. Each
of these blocks imparts a signature on the transmitted signal that is specific
to the device. For example, a DAC's input-output characteristics impose a
nonlinear relationship known as the Integral Nonlinearity (INL), which measures
the deviation between the ideal and measured output values for a
given input. Furthermore, the poles of the low-pass filters can deviate
slightly from their nominal location due to component manufacturing tolerances.
This deviation also introduces a device-specific imperfection. Additionally,
power amplifiers exhibit nonlinear behavior as input levels increase and may
introduce in-band noise or interference in adjacent frequency channels. Lastly,
the oscillators can introduce a phase imbalance in the data since there will
always be a small phase offset between them causing the resulting I and Q
channels to deviate slightly from perfect  quadrature. All of these effects
could be exploited for the purposes of device identification.

A simplified model, in complex exponential form, for an ideal signal of bandwidth $W$, transmitted by a wireless device is,
\begin{align}
\begin{aligned}
    x\left( t \right) =\left[ h_{tx} * \frac{1}{2\pi} \int_{-W/2}^{W/2} X(w) e^{jwt} dw \right] e^{jw_c t}
\label{eq:sei_model}
\end{aligned}
\end{align}
where $h_{tx}$ is the complex-valued impulse response for in-phase and quadrature components of the low-pass filters,
$X(w)$ is the spectrum of the modulated signal, $w_c$ is the carrier angular frequency, and $*$ denotes convolution.
By the time the signal reaches the receiver, it would have propagated through a wireless multipath channel.
Assuming that the channel remains stationary over the transmission time, then it can be 
modeled by a Linear Time-Invariant (LTI) system. Thus, the received signal can be modeled as,
\begin{align}
\begin{aligned}
r(t) = h_c * x(t) + n(t),
\label{eq:rx_model}
\end{aligned}
\end{align}
where $h_c$ is a complex-valued impulse response of the FIR propagation channel, and
$n(t)$ is a circularly symmetric Gaussian noise term.

For our fingerprinting problem, the training data contains isolated signal
bursts (each burst associated with a single device) in which the propagation
channels represented in train and test sets are given by $h_c^{train}$ and $h_c^{test}$ respectively.
We are interested in the case when $h_c^{train} \neq h_c^{test}$ (e.g. $h_c^{train}$
may follow a Rayleigh model while $h_c^{test}$ follows a Ricean model).
As stated previously, the performance of traditional CNNs will degrade
significantly in these scenarios due to overfitting on the channel in the training set.
Our proposed CNN architecture is resilient to these distribution changes, however, which results in improved generalization. 

%\begin{figure}
 %\centering
  
 %   \includegraphics[width=0.5\textwidth]{figures/fingerprint_envelope.png}
  
%\caption{TODO: FIX SPACING AND SIZING. For a given RF transmission specification (the black dashed lines that envelop %the solid black line), there are infinitely many specific trajectories that transmissions can take (the blue, red, and green %lines). Each trajectory is emitted from a unique device and is the basis of the fingerprinting problem.}
%\label{subfig:fingerprint_envelope}
%\end{figure}

\begin{figure*}[t!]
\centering
\includegraphics[trim=0 190 0 65, clip, width=\textwidth]{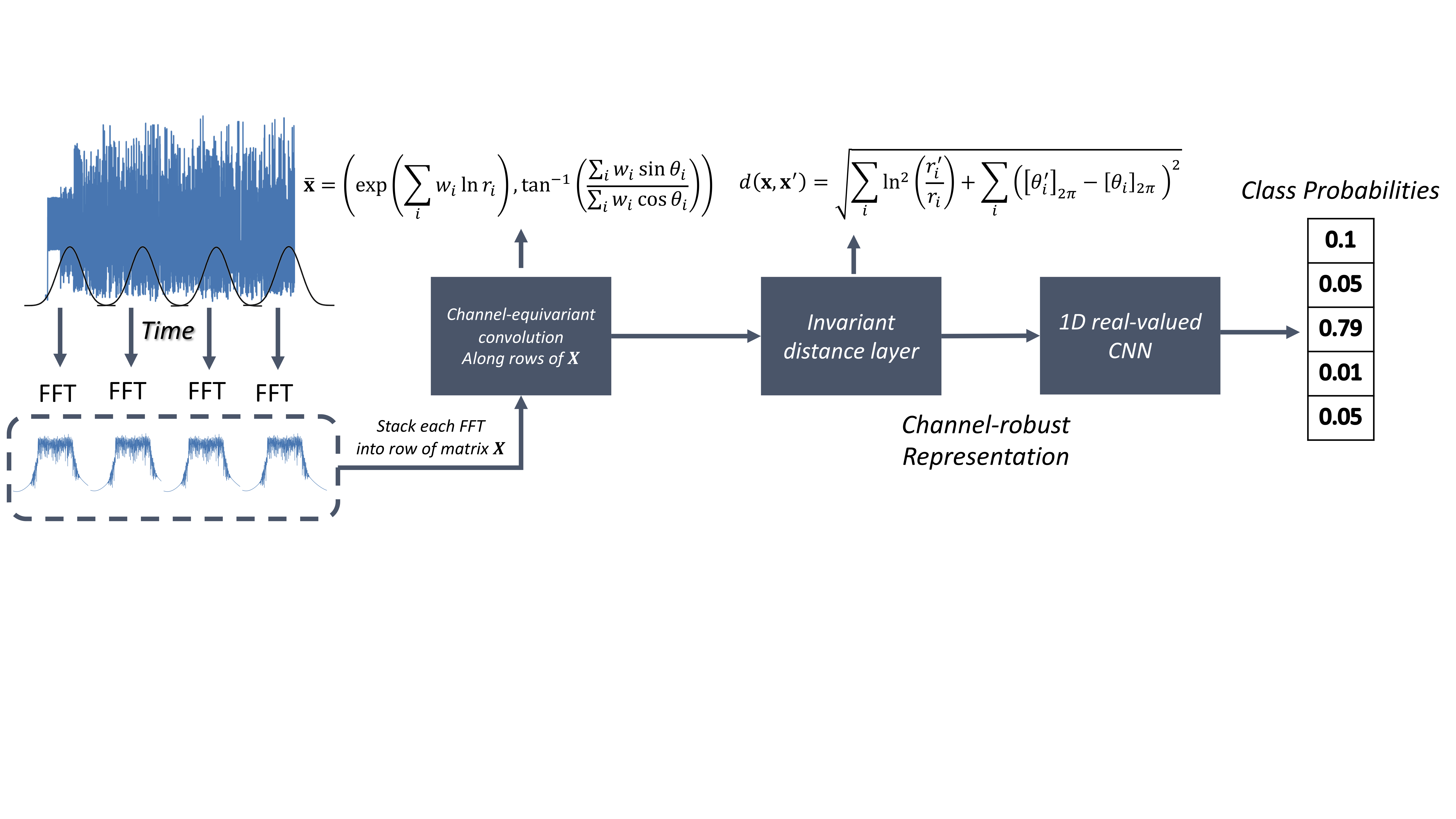}
\caption{Schematic of the ChaRRNet pipeline. An STFT is applied to the input signal using a Kaiser window.
The equivariant layer operates on the time-normalized short-time spectra.
The invariant layer computes the distance on the manifold between the output features of the equivariant layer
and their Lie-algebraic mean. The output of the invariant layer is used as input to a real-valued backbone 1D CNN for classification.}
\label{fig:arch_schematic}
\end{figure*}

\subsection{Group Theoretic Framework}
\label{subsec:group_theoretic_framework}
%In this section we present the background and framework necessary for building
%CNN layers that respect the symmetries the we wish to exploit for the problem of
%RF fingerprinting under multipath channels.

Recall the received signal model, $r(t) = x(t)*h_c + n(t)$.
We wish to extract features from $r(t)$ which are invariant to the channel.
Let $f: \mathcal{C}^N \mapsto \mathcal{C}^M$ be a learned feature extractor.
For $f$ to be invariant to the channel we must have 
$f \left( r(t) \right) = f\left(r'(t) \right)$,
where $r'(t) = x(t) + n(t)$. We call $r(t)$ a transformed version of $r'(t)$.
Of particular interest are sets of transformations that produce groups.
Consider $SO(2)$, the set of all rotations about the origin in $\mathbb{R}^2$.
Any combination of 2D rotations is also a 2D rotation, the order of rotations does not matter,
and any rotation can be undone by the inverse rotation.
These properties make $SO(2)$ an Abelian group, with matrix multiplication as the group operation.
%An Abelian group follows the following properties:
%\begin{enumerate}
%    \item Closure: $\forall a, b \in \mathbb{Z}: a \cdot b\in\mathbb{Z}$
%    \item Commutativity: $\forall a,b\in\mathbb{Z}: a \cdot b=b \cdot a$
%    \item Associativity: $\forall a,b,c\in\mathbb{Z}: (a \cdot b) \cdot c=a \cdot (b \cdot c)$
%    \item Identity: $\forall a\in\mathbb{Z}: a \cdot 0=0 \cdot a=a$, and $0$ is the identity
%    element
%    \item Inverse: $\forall a\in\mathbb{Z}: \exists b$ such that $a \cdot b=b \cdot a=0$
%\end{enumerate}
Another class of groups is Lie groups, which are continuous and
smooth groups that form a manifold. 
We shall explore the properties of a Lie
group by example with $SO(2)$. Note that for any $g,h\in SO(2)$, 
the properties of an Abelian group are satisfied
under the operation of matrix multiplication. Furthermore, for any angle $\theta$,
there is an exponential map from it to an element of the group. In this case, 
with $A = \begin{pmatrix}
     0 & -\theta\\
    \theta & 0
\end{pmatrix}$
,
$e^{A} = 
\begin{pmatrix}
\cos\theta & -\sin\theta\\
 \sin\theta & \cos\theta
 \end{pmatrix}
 $.
 Conversely, the logarithmic map is the inverse of the exponential map and
 takes an element of the group to its Lie algebra, which can be viewed as
 the tangent space to the Lie group at the identity.
 
\subsection{Groups for RF Channel Propagation}
\label{subsec:channel_propagation}
In our design of RF channel equivariant and invariant layers, we largely build upon
the work by \cite{chakraborty_surreal_2020}, which introduces the group for
complex-valued inputs as $\mathbb{R}^+\times SO(2)$ and defines operations
leveraging the wFM. However, this group models scalar multiplication.
Below, we extend this work to the multipath model and note that all operations can
be defined simply by the Lie algebra of the group.

RF signals are complex-valued, so each element of the vector can be written in
polar coordinates as $re^{i\theta}, r\in\mathbb{R}^+, \theta\in\mathbb{R}$. Note
that the radial and angular components each form independent Lie groups: the radial
component is the Abelian Lie group $\mathbb{R}^+$ under scalar
multiplication, and the polar component is the Abelian Lie group
$U(1)=\{e^{i\theta}|\theta\in\mathbb{R}/2\pi\mathbb{Z}\}$, which is isomorphic
to $SO(2)$. The logarithmic map for each component is $\ln$.
%which takes
%the radial component to $\mathbb{R}$ and the angular component to
%$\{i\theta|\theta\in\mathbb{R}/2\pi\mathbb{Z}\}$.
This Lie group $\mathbb{R}^+\times U(1)$ captures scalar
attenuation, i.e. gain in the transmission chain, and angular rotations, i.e. a
phase offset.
The simplified received signal model (\ref{eq:rx_model}), reduces the wireless channel to an LTI
system. Under this simplification, the transmitted signal, $x(t)$, is convolved with the channel
impulse response, $h_c$. By the convolution theorem,
$\mathcal{F}(h_c(t)*x(t)) = \mathcal{F}(h_c)\cdot\mathcal{F}(x)$,
where $\mathcal{F}$ is the Fourier Transform operator. 
%This means that for any signal
%hat passes through the same channel, each frequency bin is multiplied 
%by the same complex value. 
Hence, each frequency component lies on the above manifold and perturbations from the channel are
$\mathbb{R}^+\times U(1)$ at each frequency. Our RF model seeks invariance to these per-frequency perturbations.

We now define equivariant and invariant operations on this manifold.
We perform these network operations by first mapping inputs to the tangent space,
applying algebraic operations, then mapping back to the manifold. 
First, note that for any two points $r_1e^{i\theta_1}, r_2e^{i\theta_2}\in\mathbb{R}^+\times U(1)$, we can
compute means on the manifold as follows. For the radial component, we map to
the tangent space, compute a mean, and map to the manifold producing
$\exp{\frac{\ln{r_1}+\ln{r_2}}{2}}=\sqrt{r_1\cdot r_2}$, i.e. the geometric
mean. For the angular component, we obtain
$\exp{i\frac{\theta_1+\theta_2}{2}}$, which we compute via the angular component of the
directional mean, $\arctan{\frac{\sum_i{\sin\theta_i}}{\sum_i{\cos\theta_i}}}$,
as shown in Figure \ref{fig:arch_schematic}. These means are equivariant to a scaling and a rotation, respectively.
If both points are scaled by $\rho$ and rotated by $\phi$, then the means become
$\sqrt{\rho \cdot r_1 \cdot \rho \cdot r_2}=\rho\cdot\sqrt{r_1\cdot r_2}$ and
$\exp{i\frac{(\phi+\theta_1)+(\phi+\theta_2)}{2}}=\exp{i\phi}\cdot\exp{i\frac{\theta_1+\theta_2}{2}}$.
Note that this generalizes to collections of $N$ points and weighted means.

Furthermore, we can calculate distances on the manifold by mapping distances in
the tangent space with the logarithmic mapping. Then, for any two points
$r_1e^{i\theta_1}, r_2e^{i\theta_2}\in\mathbb{R}^+\times U(1)$, we shall compute
their manifold distances. For the radial component, $d_r(r_1,r_2)=|\ln{r_2} -
\ln{r_1}|=|\ln{r_2/r_1}|$. For the angular component,
$d_\theta(\exp{i\theta_1},\exp{i\theta_2})=|i([\theta_2]_{2\pi}-[\theta_1]_{2\pi})|=|[\theta_2]_{2\pi}-[\theta_1]_{2\pi}|$.
These distances are invariant to a scaling and a rotation, respectively. If both points are
scaled by $\rho$ and rotated by $\phi$, then the distances become $d_r(\rho\cdot
r_1, \rho\cdot r_2)=|\ln{(\rho\cdot r_2)/(\rho\cdot r_1)}|=|\ln{r_2/r_1}|$ and
$d_\theta(\exp{i\phi}\cdot\exp{i\theta_1},\exp{i\phi}\cdot\exp{i\theta_2})=|[\phi+\theta_2]_{2\pi}-[\phi+\theta_1]_{2\pi}|=|[\theta_2]_{2\pi}-[\theta_1]_{2\pi}|$.
If a line element is described by $ds^2 = dr^2 + d\theta^2$, then we obtain
a total distance of $d(r_1e^{i\theta_1}, r_2e^{i\theta_2})=\sqrt{\ln^2(r_2/r_1)
+([\theta_2]_{2\pi}-[\theta_1]_{2\pi})^2}$.

%\begin{figure}
%\includegraphics[width=0.5\textwidth]{example-image-a}
%\caption{Possibly schematic of the complex plane to half-cylinder manifold for 1-dimension?}
%\end{figure}

\subsection{Network Architecture}
\label{subsec:architecture}
Leveraging the Abelian Lie groups defined above, we can design neural
network layers that are equivariant or invariant to multipath channels.

\subsubsection{Equivariant Layers}

Given that means taken with respect to a manifold are equivariant per frequency bin,
suppose that we have $N$ input windows of signals that have all gone through the same channel.
Then, if we compute the above means per frequency bin,
they are equivariant to the channel's frequency response for each
bin. So, equivariant layers are generalized means that perform convolution by
striding the mean for the Abelian Lie group with learned weights along a
tensor's axis that contains different windows in the frequency domain.

\subsubsection{Invariant Layers}
Similarly, leveraging the fact that distances on the manifold are
invariant per frequency bin, we can design an invariant layer by
computing the distance from the equivariant means to their respective inputs.
Since the inputs all experienced the same frequency response and the mean
was equivariant, the same frequency response is present for both the input and
the mean, and the distance produces a value invariant to the frequency response.
Thus, invariant layers first perform a learned convolution as done in the 
equivariant layer before computing an elementwise distance from each input
window of the mean to the mean per frequency bin.

%\caption{Output of the equivariant layer. Top row: radial component.
%Bottom row: angular component. Let $H_c, X$ be the Fourier transform
%of the channel, $h_c$, and input signal, $x$, respectively.
%The left column shows $f_{equiv} \left(H_c\cdot X \right)$.
%The middle column shows $H_c \cdot f_{equiv} \left(X\right)$.
%The right column shows $\Big| f_{equiv} \left(H_c\cdot X \right) -
%H_c \cdot f_{equiv} \left(X\right) \Big|$, 
%verifying that the layer is indeed equivariant to the channel impulse response.
%}
%\label{fig:equiv_layer}
%\end{figure}

\begin{figure*}[t!]
\centering
\scalebox{0.9}{
\begin{tabular}{cc}

	\includegraphics[width=\columnwidth]{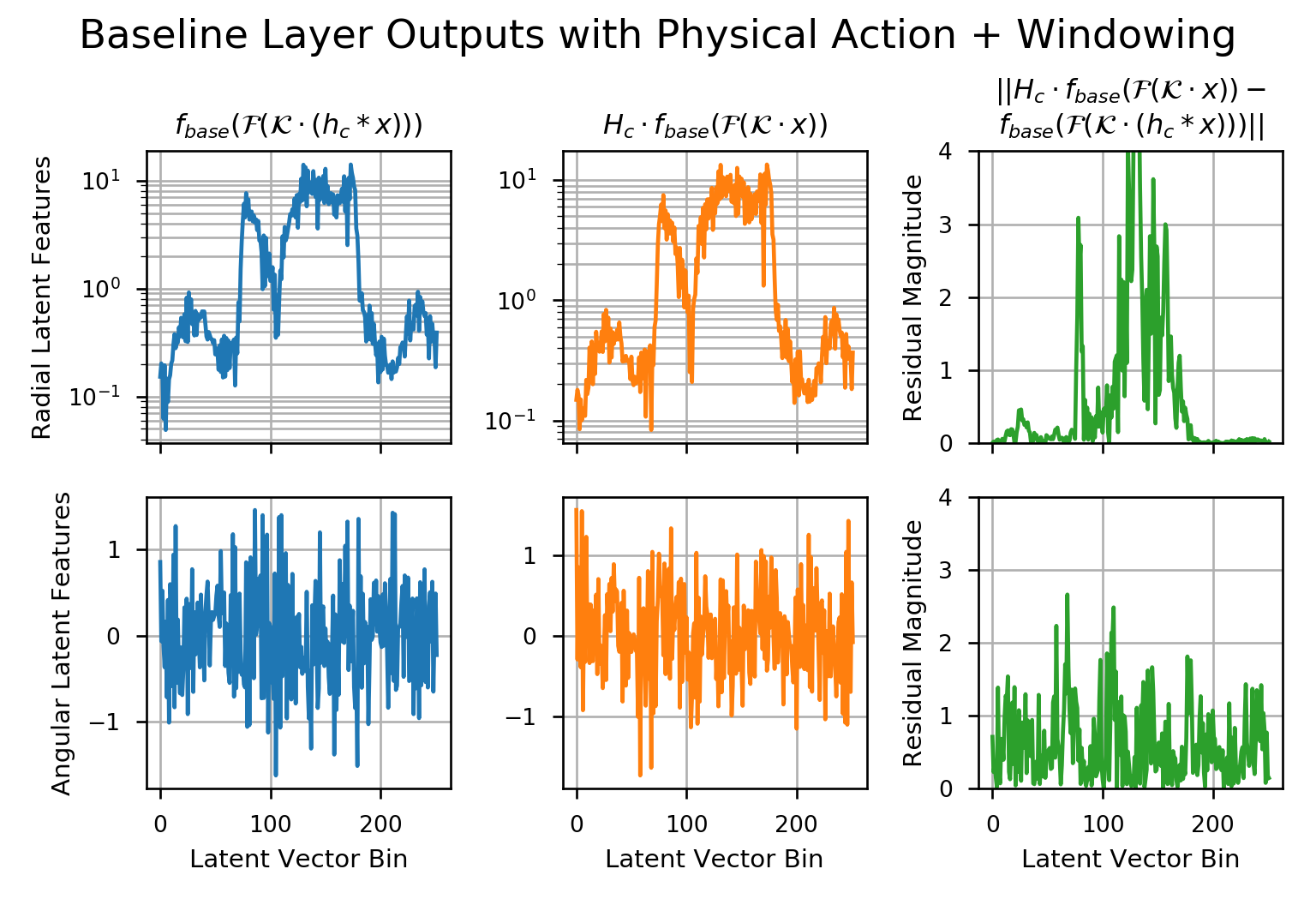}
	 &
	
	\includegraphics[width=\columnwidth]{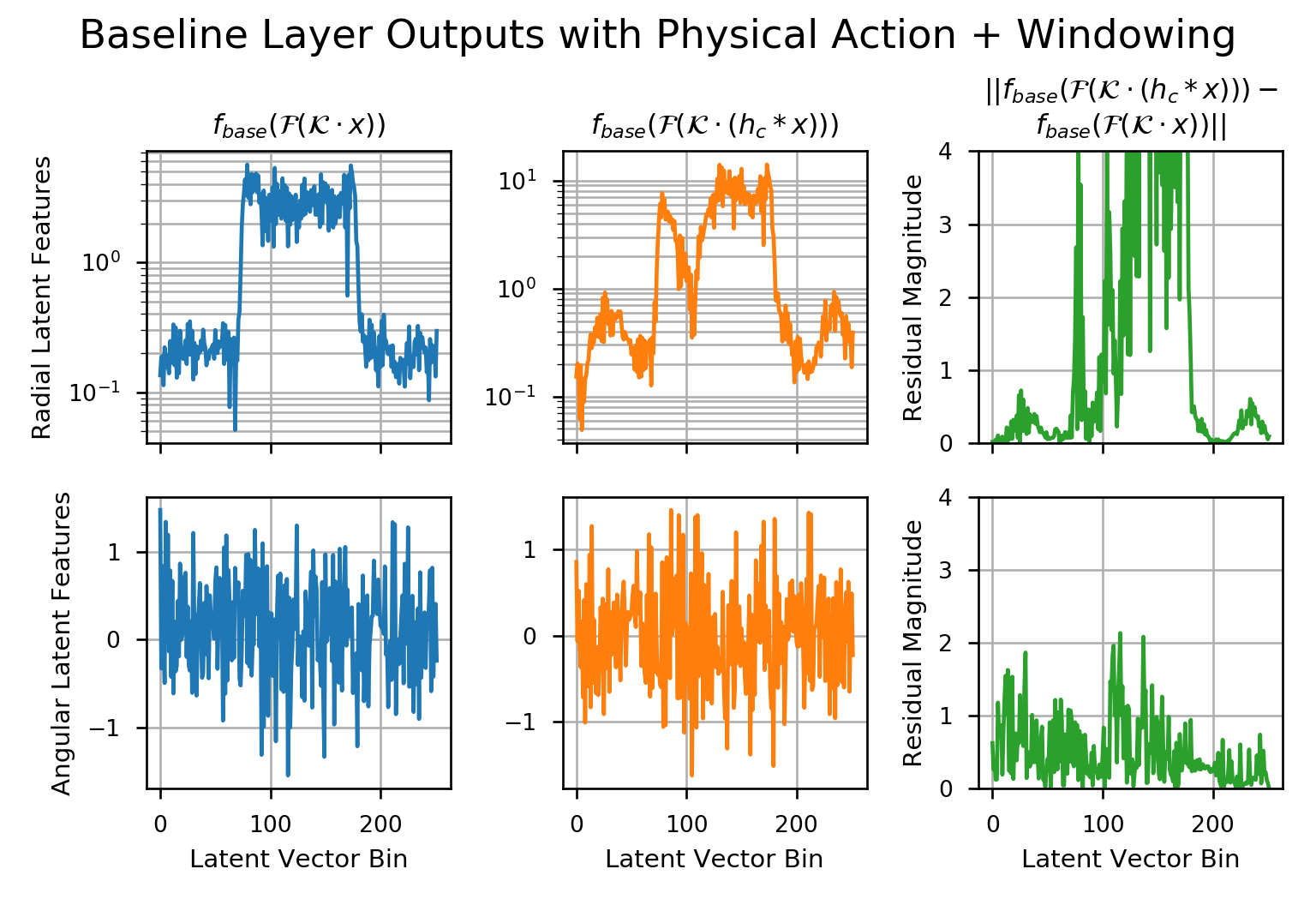}
	 \\
	
   	\includegraphics[width=\columnwidth]{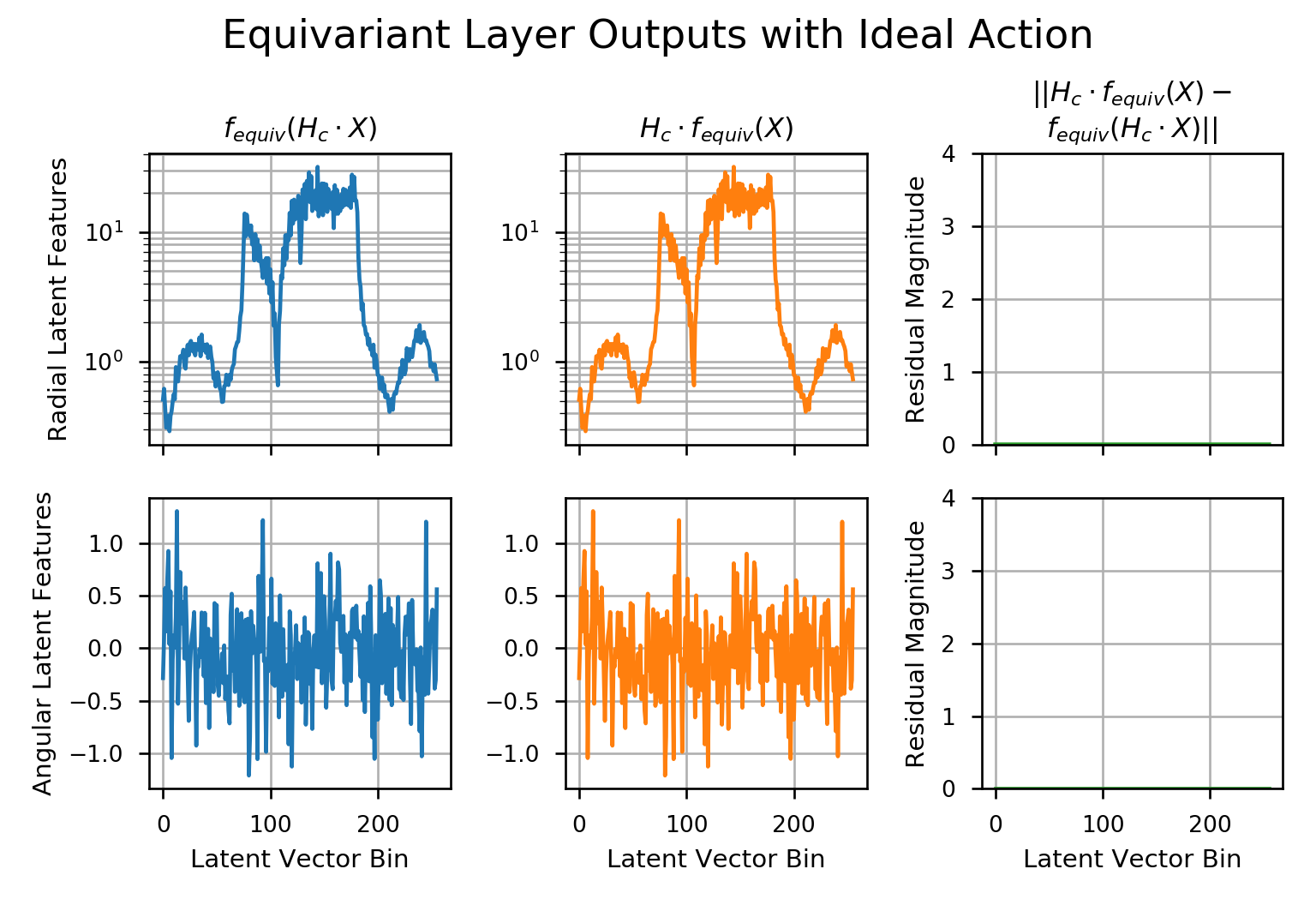}
   	 &
  
	\includegraphics[width=\columnwidth]{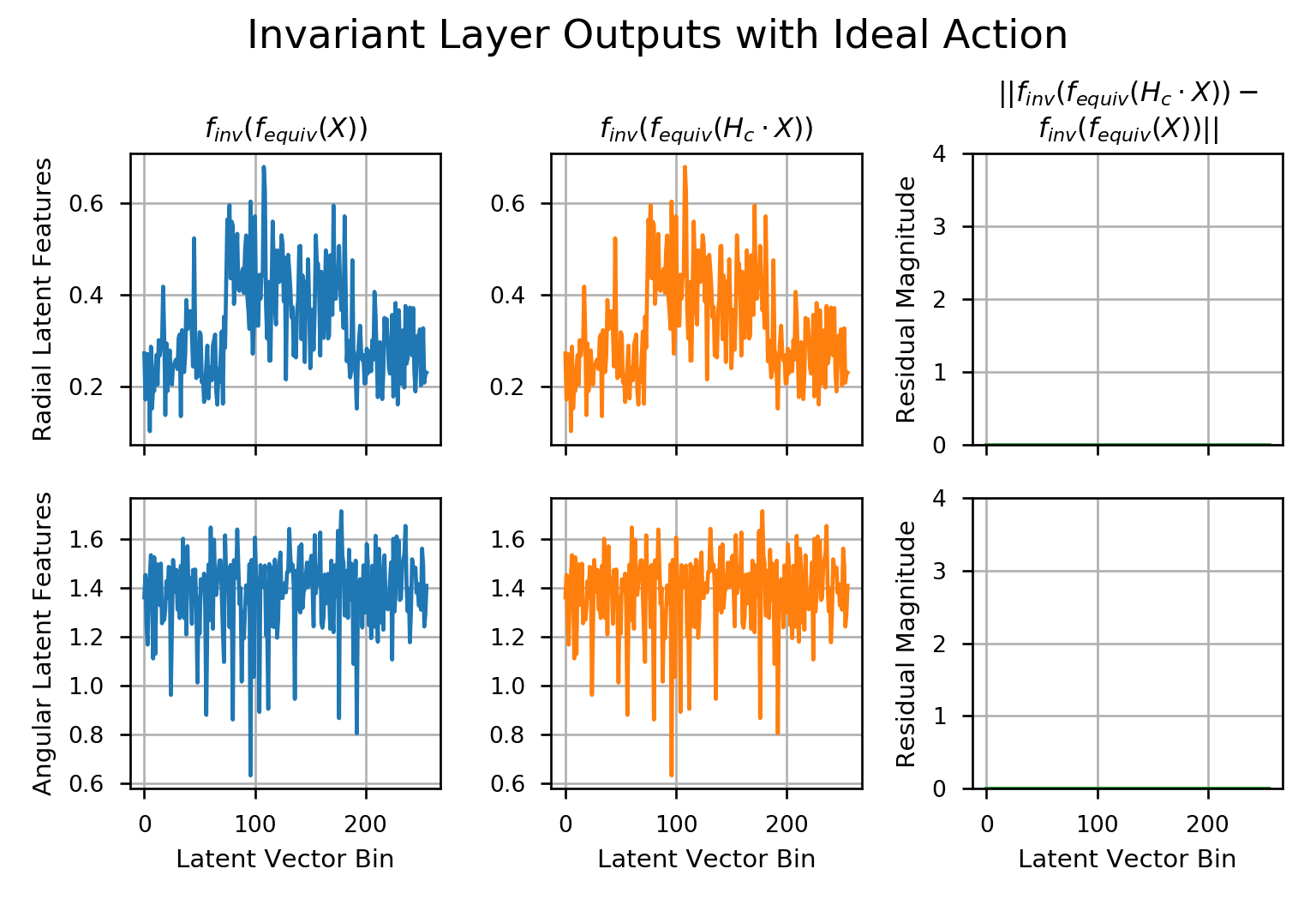}
	 \\

	\includegraphics[width=\columnwidth]{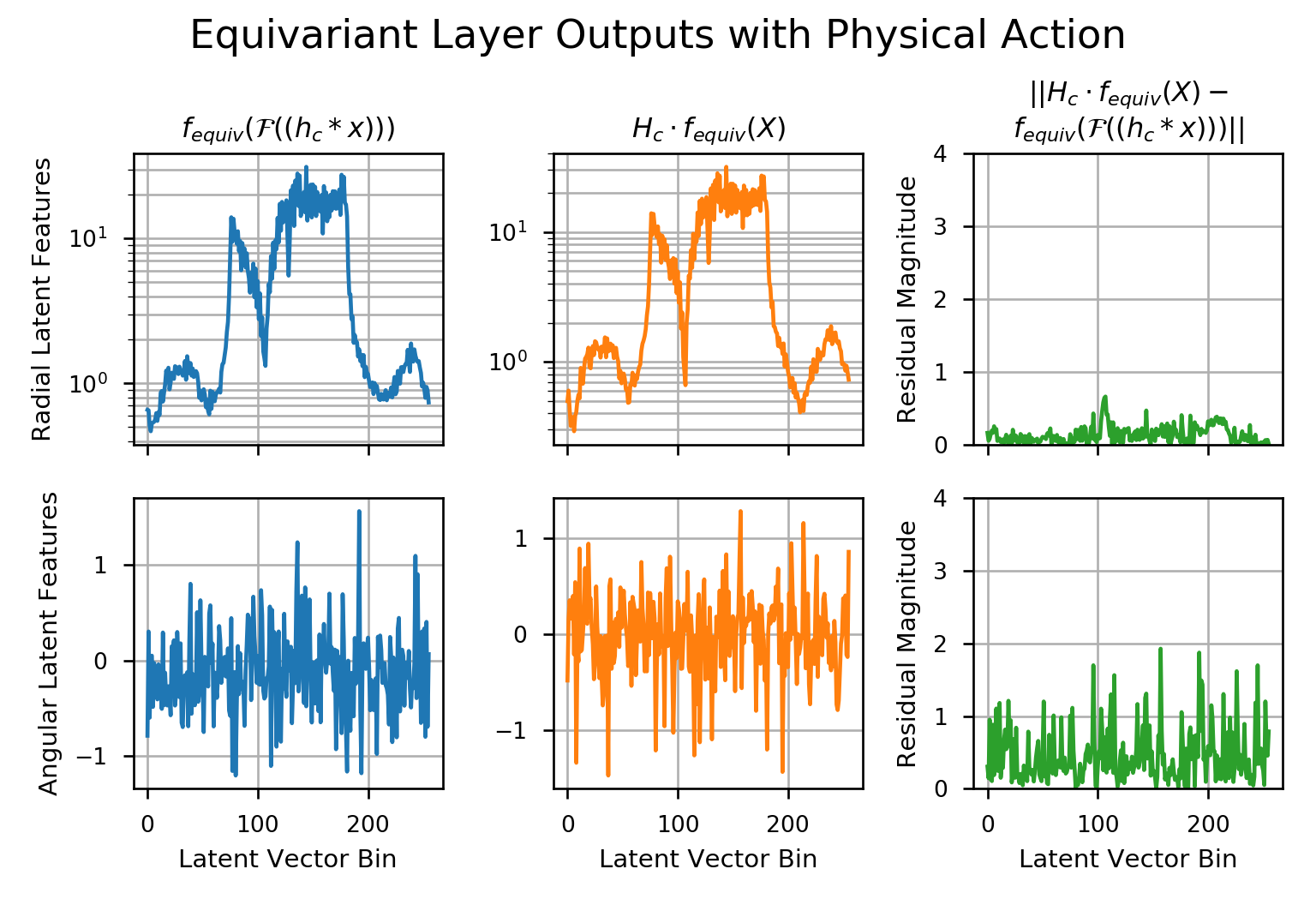}
	 &
	
	\includegraphics[width=\columnwidth]{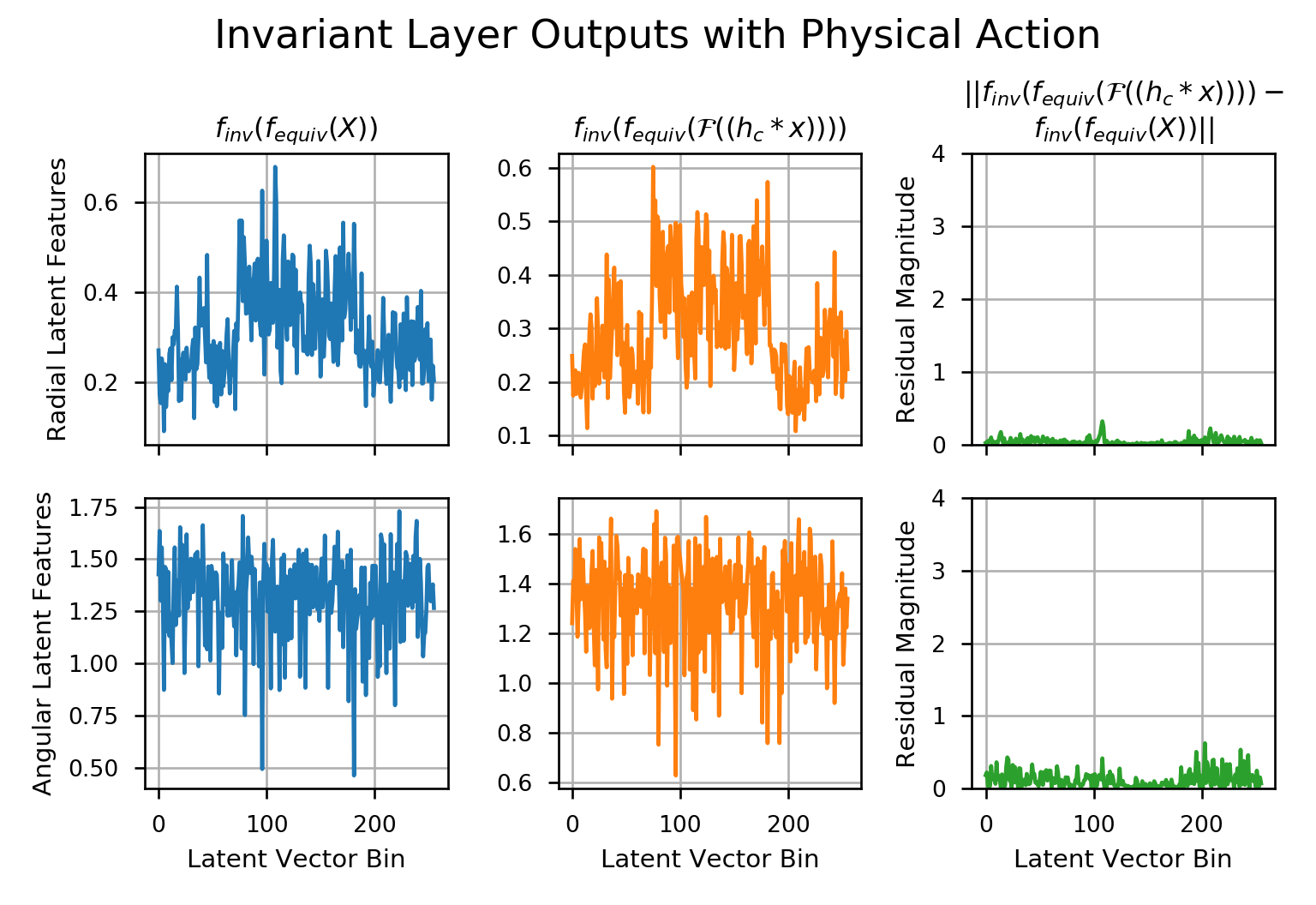}\\
	
	\includegraphics[width=\columnwidth]{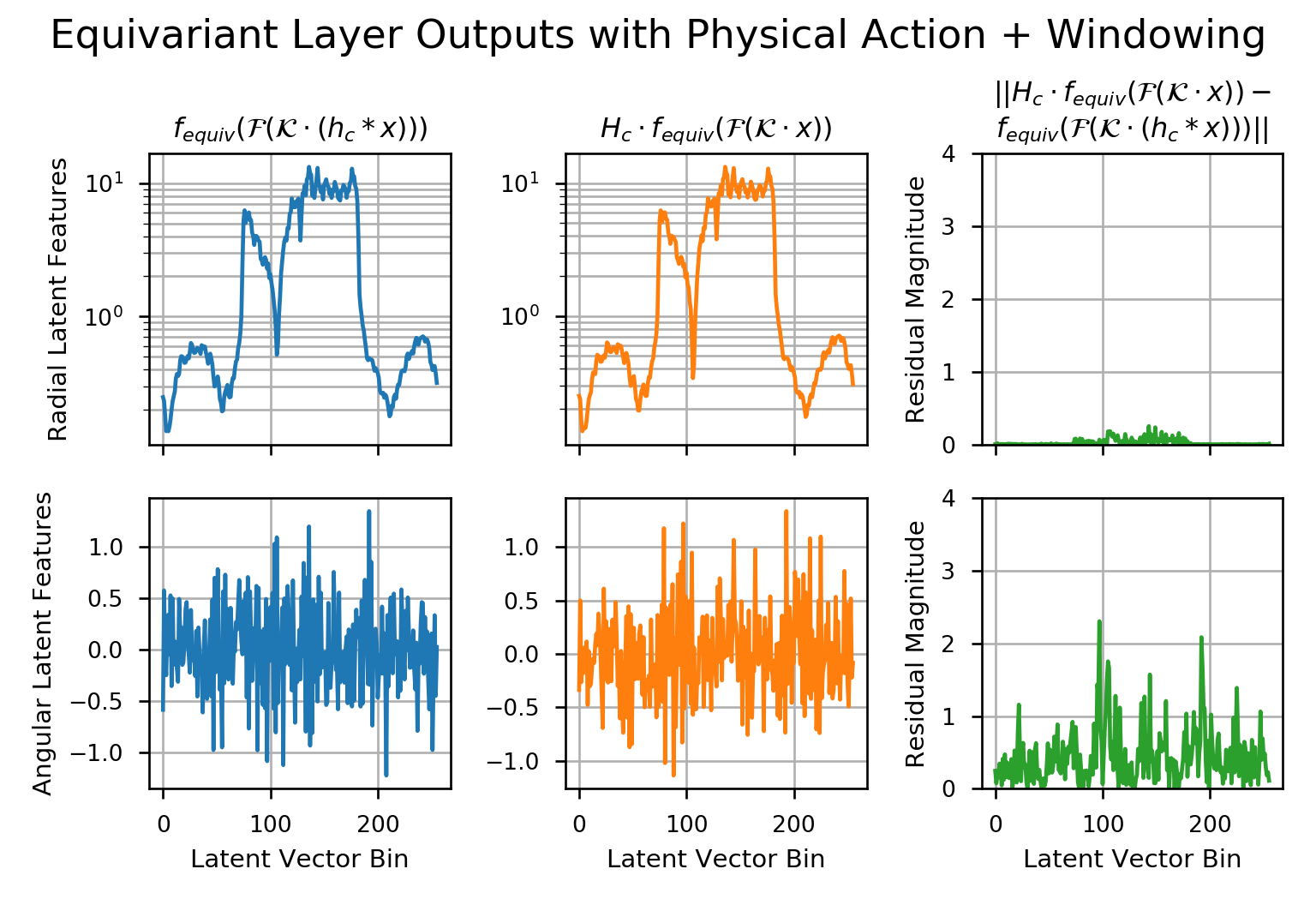}\
	 &
	
	\includegraphics[width=\columnwidth]{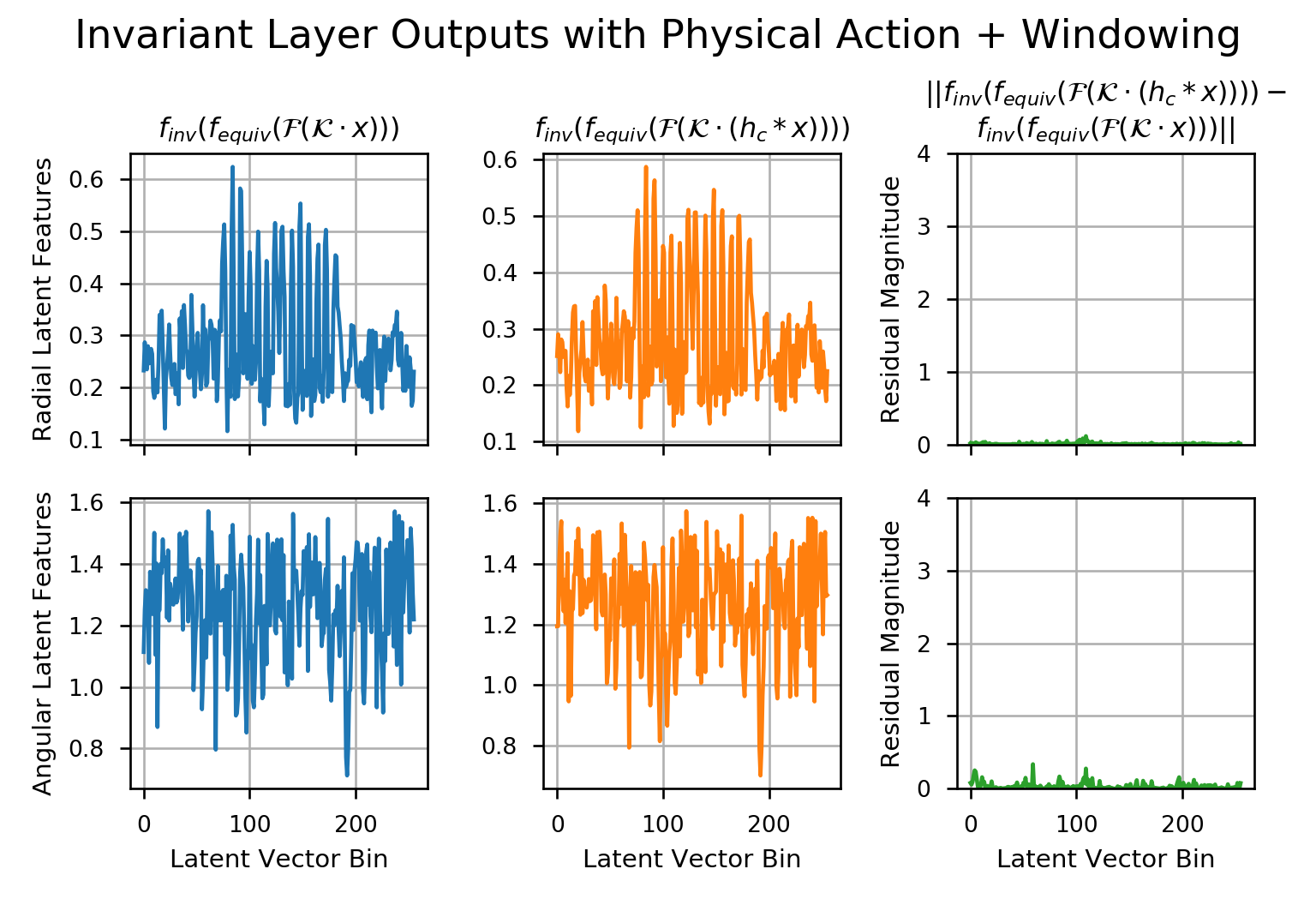}
	
\end{tabular}}

\caption{Output of different CNN layers. Here, we test the layer's equivariance and invariance properties.
The first row shows the results of the Baseline complex-valued convolutional layer.
The left column tests for layer equivariance, the right column tests for layer invariance.
The ideal action applies the channel independently to each STFT window by element-wise 
multiplication in the frequency domain. Our layers are mathematically equivariant and invariant to this ideal action.
The physical action applies the channel by convolving the input signal with the channel's impulse response, while
the physical action + windowing applies the physical action and helps to mitigate the boundary effects discussed in
Section~\ref{subsec:architecture}. These results show that our layers compute more stable representations
relative to a traditional complex-valued convolutional layer under a multipath channel.
}
\label{fig:equivariance_invariance_tests}

%Output of the equivariant layer. Top row: radial component.
%Bottom row: angular component. Let $H_c, X$ be the Fourier transform
%of the channel, $h_c$, and input signal, $x$, respectively.
%The left column shows $f_{equiv} \left(H_c\cdot X \right)$.
%The middle column shows $H_c \cdot f_{equiv} \left(X\right)$.
%The right column shows $\Big| f_{equiv} \left(H_c\cdot X \right) -
%H_c \cdot f_{equiv} \left(X\right) \Big|$, 
%verifying that the layer is indeed equivariant to the channel impulse response.
%}

\end{figure*}

\subsubsection{Complete Model}
With equivariant and invariant layers, now we can stack them to design a model
that is inherently robust to channel effects. A general, high-level, 
description of this model is depicted in Figure~\ref{fig:arch_schematic}.
We compute the Short Time Fourier Transform (STFT) of the input signal burst using a
Kaiser window. Then, we apply an equivariant layer by computing learned weighted 
sums with the Lie algebra's mean. Finally, we apply an invariant layer,
which outputs real-valued latent representations. The invariant vectors are input to a 
real-valued backbone 1D CNN for classification.

Once we compute the STFT, we require each window to have been convolved with the
channel independently of each other. In reality, the first $L-1$
samples of each window have contribution
from the previous window, and the last $L-1$ samples where the convolution
dies past the edge of the samples in a given window are not present.
These boundary effects break the theoretical equivariance and invariance of the the algebraic layers. 
However, under the assumption that the length of the channel impulse response is small relative to 
the length of a short-time window, this boundary effect can be 
mitigated by applying a window function such as a Kaiser window when computing the STFT.
The difference between ChaRRNet's equivariance and invariance to the
mathematical ideal of the transformation and the
physical reality along with the Kaiser window's effect are shown in Figure~\ref{fig:equivariance_invariance_tests}.

\section{Results}
\label{sec:results}

%For the mathematical model at basis of FIRNet's design, we formulated the
%problem in the
%frequency domain. However, as a realistic signal propagates, the channel effects
%accumulate in the time domain. Next, we verify the invariance of
%ChaRRNet in the frequency domain and examine the model's robustness on 
%synthetic and real data and compare it to a non-invariant
%complex-valued neural network for RF fingerprinting.

\subsection{Simulated Transmitter Fingerprinting}

We now evaluate ChaRRNet against 802.11g signals generated using GNURadio, where the
device fingerprints are imparted using an IIR filter, a standard model for WiFi signals.
We set the population size to 65 devices. Each burst passes through a simulated channel
with a specified number of reflectors. The attenuation of each reflector is drawn from
$\mathcal{U}[-15dB, -5dB]$, and its nominal position is slightly randomized.
Each burst is convolved with an impulse response drawn randomly from the reflector geometry.
We use a larger and deeper version of the complex-valued convolutional network of \cite{exp_fingerprinting_2019} as a baseline. 
We train both models using two datasets:
(1) pristine, i.e. no channel applied to bursts, and (2) nLOS 200,
where realizations from a channel geometry without a line-of-sight and 200 reflectors
are applied to the signal bursts.
Both models are trained with additive white Gaussian noise (AWGN) and center frequency offset (CFO) data augmentation.
The Baseline model has 5M trainable parameters, and the ChaRRNet model has 4M trainable parameters.
The results are presented in Table \ref{table:synth_results}.

%\begin{figure}[t!]
%\centering
%\begin{subfigure}
%	\centering
%	\includegraphics[width=\columnwidth]{figures/baseline_layer-physical-kaiser-inv.png}
%	\label{fig:y equals x}
%\end{subfigure}
%\begin{subfigure}
%	\centering
%	\includegraphics[width=\columnwidth]{figures/invar_layer.png}
%	\label{fig:y equals x}
%\end{subfigure}
%\begin{subfigure}
%	\centering
%	\includegraphics[width=\columnwidth]{figures/invar_layer-physical.png}
%	\label{fig:y equals x}
%\end{subfigure}
%\begin{subfigure}
%	\centering
%	\includegraphics[width=\columnwidth]{figures/invar_layer-physical-kaiser.png}
%	\label{fig:y equals x}
%\end{subfigure}
%\caption{Output of the invariant layer. Top row: radial component. Bottom row: angular component. Let $H_c, X$ be the Fourier transform
%of the channel, $h_c$, and input signal, $x$, respectively.
%The left column shows $f_{inv} \left(f_{equiv} \left( X\right)\right)$.
%The middle column shows 
%$f_{inv} \left(f_{equiv}\left(H_c\cdot X \right)\right)$. 
%The right column shows $\Big| f_{inv} \left(f_{equiv} \left( X\right)\right) -
%f_{inv} \left(f_{equiv}\left(H_c\cdot X \right)\right) \Big|$, 
%verifying that the layer is indeed invariant to the channel impulse response.
%}
%\label{fig:invar_layer}
%\end{figure}

\begin{table}[t!]
	\centering
	
	\caption{Results for training Baseline and ChaRRNet models on both pristine WiFi signals
             and signals that have gone through multipath with 200 simulated reflectors.
             We evaluate on test sets with multipath configurations varying
             from 0 to 400 reflectors with and without line of sight.}
             
	\begin{tabular}{|c || c| c || c |c|}
		\hline
		 \multirow{2}{4em}{Test Set} & \multicolumn{2}{c||}{Baseline Trained on:} & \multicolumn{2}{c|}{ChaRRNet Trained on:} \\
		 & Pristine & LOS200 & Pristine & LOS200 \\ 
		\hline
		Pristine & 80.4 & 23.5 & \textbf{85.5} & \textbf{24.1} \\
		nLOS 10  & 11.7 & 76.7 & \textbf{20.9} & \textbf{83.9} \\
		LOS 10   & 5.5 & 76.5 & \textbf{23.6} & \textbf{84.0} \\
		nLOS 30  & 11.8 & 78.4 & \textbf{26.8} & \textbf{84.5} \\
		LOS 100  & 5.4 & 70.5 & \textbf{25.6} & \textbf{83.0} \\
		nLOS 400 & 10.8 & 78.6 & \textbf{25.4} & \textbf{85.1} \\
		LOS 400  & 11.5 & 79.3 & \textbf{25.4} & \textbf{85.3} \\
		\hline
	\end{tabular}
	
	\label{table:synth_results}
\end{table}
We find that the ChaRRNet model obtains stable generalization and higher accuracy
across a wide range of simulated multipath channels.
Also note that training with the simulated channel improves generalization to other channels, a 
result that motivates the use of channel data augmentation during training.

%We perform a UMAP \textbf{REFERENCE} dimensionality reduction to visualize
%the difference in learned latent features between the the two models.
%In these plots \textbf{SHOW UMAP PLOTS OF CHANNEL TESTS}, we note that the
%per-class clusters for FIRnet are significantly tighter, implying better
%class understanding in spite of operating in unknown channel environments.

\subsection{Real Transmitter Fingerprinting}
Next, we examine model performance on real signals collected in the wild.
We use two datasets for evaluating the model performance:
\begin{enumerate}
    \item The training and testing sets are comprised of outdoor-collected WiFi signals
          from separate days. Thus, we have unique channel conditions during training
          that are separate from those seen during testing.
    \item The training consists of two days of outdoor-collected WiFi signals while
          the test-set still contains a single separate day. We expect this to
          be an easier test case, as the train data contains a diversity of channels
          across days.
\end{enumerate}
Each day ostensibly has a different set of channel conditions.
There are 50 devices in the first dataset, while there are 65 devices in the second dataset.
For all experiments, we train with data augmentation using simulated AWGN and CFO. We also
augment the training set by applying multipath channels drawn randomly from Rayleigh and Ricean
models. 
In both cases, we see that the ChaRRNet vastly outperforms the baseline complex-valued
convolutional architecture, substantiating the claim that invariances lead to
greater generalization across unseen domains. Furthermore, we note that the case of
training on two separate days in Table \ref{table:channel_results} provides a significant boost in performance, as
it further helps the model to correctly calibrate its weights to the existence of channel artifacts.

\begin{table}[t!]
	\centering
	
	\caption{Comparing ChaRRNet and Baseline top-1 accuracy on real signals with a
	channel distribution shift from train to test.}
		
	\begin{tabular}{|>{\centering\arraybackslash}m{.42\linewidth}|>{\centering\arraybackslash}m{.16\linewidth}|>{\centering\arraybackslash}m{.1\linewidth}|>{\centering\arraybackslash}m{.12\linewidth}|>{\centering\arraybackslash}m{.1\linewidth}}
		\hline
		Dataset Description & Average Train Signals per Device & Baseline & ChaRRNet \\
		\hline
		Train 1 Day, Test Holdout Day & 100 & 9.4 & \textbf{44.8} \\
		Train 2 Days, Test Holdout Day & 218 & 25.6 & \textbf{65.5} \\
		\hline
	\end{tabular}
	
	\label{table:channel_results}
\end{table}

Finally, we compare ChaRRNet to the Baseline model across a broad range of signal fingerprinting tasks in Table \ref{table:more_expansive_results} that contain
mixes of WiFi and ADS-B signals.

\begin{table}[t!]
	\centering
	
	\caption{More extensive comparisons of top-1 accuracy between ChaRRNet and the baseline model.
             Note that ChaRRNet obtains the best performance on the 10K device dataset despite
             having 20\% fewer parameters. Note that on average each device contains $22$ training samples.
             The large difference in performance on this large population dataset can be attributed to ChaRRNet being
             more data efficient due to the incorporation of the multipath propagation inductive bias.}
             
	\begin{tabular}{|>{\centering\arraybackslash}m{.42\linewidth}|>{\centering\arraybackslash}m{.16\linewidth}|>{\centering\arraybackslash}m{.1\linewidth}|>{\centering\arraybackslash}m{.12\linewidth}|>{\centering\arraybackslash}m{.1\linewidth}}
		\hline
		Dataset Description & Average Train Signals per Device & Baseline & ChaRRNet \\
		\hline
		100 Devices & 218 & \textbf{93.2} & 92.1 \\
		1000 Devices & 218 & 81.5 & \textbf{94.0} \\
		10000 Devices & 22 & 47.1 & \textbf{77.6} \\
		Low Train Size & 56 & 82.3 & \textbf{82.7} \\
		Medium Train Size & 280 & \textbf{92.6} & 92.5 \\
		Large Train Size & 447 & 93.8 & \textbf{95.1} \\
		Train/Test Same Indoor Day & 800 & 92.3 & \textbf{92.7} \\
		Train/Test Same Outdoor Day & 800 & \textbf{83.5} & 80.2 \\
		Train/Test on Mix of Days & 800 & 88.5 & \textbf{88.9} \\
		Bitwise Identical & 8953 & 96.7 & \textbf{98.6} \\
		\hline
	\end{tabular}
	
	\label{table:more_expansive_results}
\end{table}

\section{Conclusion}
\label{sec:conclusion}

In the ChaRRNet model, we leverage Lie groups to define novel convolutional
layers that are robust to multipath degradation of signals in the physical model
of wireless signal propagation. We have shown
that including this domain bias in the model's operations drastically increases model
generalization to multipath channel environments not present at train time.
Furthermore, we have shown the windowing processing necessary to bring
physical wireless signals with boundary effects closer to the idealized conditions
assumed by the model. We hope that this new set of convolutional layers opens a
new space of optimizing NNs for operational tasks in the RF domain while 
easing the burden of data collection.

\section*{Acknowledgment}
We are grateful to our co-workers--Camila Ramirez and Cass Dalton--for valuable
mathematical discussion and to Tom Rondeau, John Davies, Esko Jaska, and Paul Tilghman
for valuable feedback.

\bibliographystyle{ieeetran}
\bibliography{./bib/equivariant_nns,./bib/rf_fingerprinting,./bib/system2}

\end{document}